\documentclass{ws-ijmpd}
\usepackage[super,compress]{cite}
\begin{document}

\markboth{I.F. Ranea-Sandoval and H. Vucetich}
{Scalar Resonances in Axially Symmetric Spacetimes}

%
\catchline{}{}{}{}{}
%

\title{SCALAR RESONANCES IN AXIALLY SYMMETRIC SPACETIMES}

\author{IGNACIO F. RANEA-SANDOVAL\footnote{iranea@fcaglp.unlp.edu.ar Postdoctoral Fellow of CONICET}} \author{H\'ECTOR VUCETICH}

\address{Grupo de Gravitaci\'on, Astrof\'{\i}sica y Cosmolog\'{\i}a, Facultad de Ciencias Astron\'omicas y Geof\'{\i}sicas, Universidad Nacional de La Plata. Paseo del Bosque S/N 1900. La Plata, Argentina
}

\maketitle

\begin{history}
\received{}
\revised{}
\end{history}

\begin{abstract}
We study properties of resonant solutions to the
scalar wave equation in several axially symmetric spacetimes. We
prove that  non-axial
resonant modes do not exist neither in the Lanczos dust cylinder, the $(2+1)$ extreme BTZ
spacetime nor in a class of simple rotating wormhole solutions.  
Moreover, we find unstable solutions to the wave equation in the
Lanczos dust cylinder and in the $r^2 <0$ region of the extreme
$(2+1)$ BTZ spacetime, two solutions that possess closed timelike
curves. Similarities with previous results obtained for the Kerr spacetime are
explored. 
\end{abstract}


 \ccode{PACS numbers: 04.20.-q, 04.70.-s}

\section{Introduction}

Theorems regarding uniqueness of the solution to the scalar wave
equation in a particular case of non-globally hyperbolic spacetime were proven in Ref.~\refcite{bache}. Moreover, it was shown that, when they exist, resonant solutions to the scalar wave
equation (i.e. exponentially growing in time) must preserve the axial symmetry of the background
spacetime. The spacetime studied in Ref.~\refcite{bache} presents a chronology violating region
in which it is possible to connect any pair of events of the manifold via a timelike future pointing
curve. As usual, the existence of
this region is related to the change of the causal nature of the axial
Killing vector field from spacelike to timelike. An
extensive review of spacetimes that admit closed timelike curves (CTC's) and its relation with
causal violation is presented in Ref.~\refcite{Lobo}. 

The existence of infinitely many axially symmetric exponentially
growing with time solutions to Teukolsky's master equation was proven,
for scalar, neutrino,
electromagnetic and gravitational fields, in Ref.~\refcite{doglersv} and
Ref.~\refcite{doglers}. A relation between the existence of such unstable modes and of CTC's was proposed in order to explain the reasons of this unstable nature in Ref.~\refcite{doglers}.  

In this work we extend the results obtained in 
Ref.~\refcite{bache} to other axially symmetric, stationary spacetimes. In order to shed some light into the proposed link between the existence of CTC's and of unstable modes, we also study the linear stability against scalar perturbations of other spacetimes presenting CTC's. 

This paper is organized as follows. In Section II we extend the
results obtained in Ref.~\refcite{bache} to a series of physically
interesting spacetimes: Lanczos dust cylinder, ($2+1$) extreme BTZ
solution and two solutions of rotating wormholes. For these
spacetimes with prove that for scalar perturbations the only possible
resonances are axial. In Section III we present results that reinforce the relation between the existence of CTC's and of unstable solutions to the scalar wave equation, as was suggested in Ref.~\refcite{doglers}. Finally, we
present some general conclusions in Section IV.

\section{Non-axial resonances in axially symmetric spacetimes}

In this Section we analyze the existence of non-axial
resonant  modes for the scalar wave equation in a series of axially symmetric
and stationary spacetimes.

\subsection{The Lanczos dust cylinder}

In 1924, Cornelius Lanczos discovered an exact solution to the
non-vacuum Einstein field equations \cite{vStock1}. This solution was
rediscovered and generalized by Willem Jacob van Stockum in 1937 \cite{vStock2}.  

In this solution, the gravitational field is generated by a
cylindrical distribution of rigidly rotating dust. The density of this
perfect fluid increases with the distance to the axis of rotation
therefore discarding, almost entirely, its physical relevance.  

In cylindrical coordinates $[t \in \mathbb{R},\rho > 0,z \in
\mathbb{R},-\pi < \phi < \pi]$ the line element for this spacetime
reads: 
\begin{equation}
ds_L^2 = -dt^2 - 2\alpha\rho ^2 dt d\phi + e^{-2\alpha \rho} [d\rho ^2 + dz^2] + (\rho ^2 - \alpha^2 \rho ^4)d\phi ^2. 
\end{equation}

This spacetime is stationary, invariant under translation along the
cylinder's axis and also rotations about it. The parameter $\alpha$ fully determines the geometry and can be interpreted as the
magnitude of the vorticity vector at the rotation axis. 

As van Stockum first noticed, this spacetime admits CTC's in the
region \hbox{$\rho > 1/\alpha$}, where the light cones become tangent
to the constant $t$ planes (also notice that in this region $g_{\phi
  \phi} < 0$ and the nature of the coordinate $\phi$ changes).

We are interested in studying the scalar wave equation that, for this
spacetime, reads as: 
\begin{eqnarray} \label{Box-phi}
\Box \Phi &=& {\frac{{{e}^{2\,a\rho}}}{\rho} {\frac{{\partial }}{\partial \rho}\Phi
 \left( t,\rho,z,\phi \right) }{}}+{\frac{1}{\rho ^2} {{\frac {\partial ^{2}}{
\partial {\phi}^{2}}}\Phi \left( t,\rho,z,\phi \right) }}-
2\,\alpha{\frac {\partial ^{2}}{\partial t\partial \phi}}\Phi \left( t,\rho
,z,\phi \right) + \\
&&+ {{e}^{2\,\alpha\rho}}\left[{\frac {\partial ^{2}}{\partial {
\rho}^{2}}}\Phi \left( t,\rho,z,\phi \right) + {\frac {\partial ^{2}}{
\partial {z}^{2}}}\Phi \left( t,\rho,z,\phi \right)\right] + \left( {\alpha}^{2}{\rho}^{2
}-1 \right) {\frac {\partial ^{2}}{\partial {t}^{2}}}\Phi \left( t,
\rho,z,\phi \right) = 0 . \nonumber
\end{eqnarray}

We use some of the spacetime symmetries to factorize the test field
$\Phi(t,\rho,z,\phi) = e^{im \phi + \lambda t}f(\rho,z)$. As a result
we get (omitting the common factor $e^{im\phi + \lambda t}$) that
(\ref{Box-phi}) reduces to: 
\begin{eqnarray}  \label{Box-mode}
\frac{1}{\rho}\frac{\partial}{\partial \rho} \left(\rho \frac{\partial f(\rho , z)}{\partial \rho}\right)+{\frac {\partial ^{2}f \left( \rho,z \right)}{\partial {z}^{2}}} + 
\left[ \gamma (r) -2\,i\alpha m
\lambda\right]  {{e}^{-2\,\alpha \rho}
}f \left( \rho,z \right) = 0, 
\end{eqnarray}

\noindent where:
\begin{equation}
\gamma (r) = \left( {\alpha}^{2}{\rho}^{2}-1 \right) {\lambda}^{2}-{\frac {{m}^{2}}{{\rho}^{2}}} .
\end{equation}

It can be proven 
that equation (\ref{Box-mode})
changes its nature from hyperbolic to elliptic at $\alpha ^2 \rho ^2
=1$, the region in which the CTC's appear. This is also true in the case of the Kerr spacetime (for details see the discussion in Ref.~\refcite{doglers}).

Using the remaining symmetry we can write $f(\rho , z) = e^{iLz}f_{\rm L}(\rho)$ in (\ref{Box-mode}) to get:
\begin{eqnarray} \label{edo-Lanczos}
{\frac {d^{2}f_{\rm L}(\rho)}{d{\rho}^{2}}} + {\frac{1}{\rho} {{\frac {df_{\rm L}(\rho)}{d\rho}}}} + \left(e^{-2\alpha\rho} \left[ \gamma (r)-2\,i\alpha m
\lambda\right] - k_z ^2 \right)f_{\rm L}(\rho) = 0.
\end{eqnarray}

The well-behaved solution at the symmetry axis (for $m>0$ a similar
expression exists for negative values of $m$) and the non-divergent
asymptotic one are: 
\begin{equation} \label{ax-asy-sols}
f_{\rm L}(\rho) \sim \rho ^m \left[1 + \mathcal{O}(\rho ^2)\right] \:\:  \rho \to 0, \qquad f_{\rm L}(\rho) \sim {\rm K}(0,k_z \rho) \:\:  \rho \to \infty,
\end{equation} 

\noindent where ${\rm K}(\beta ,x)$ is the asymptotically well behaved
Bessel function of the second kind. 

Taking the wave equation (\ref{edo-Lanczos}), multiplying it by $\rho
f^*(\rho)$ and integrating the resulting expression in the $\rho \ge 0$ semi-axis,
we arrive to the following relationship: 
\begin{equation} \label{lanczos-cond}
0 = \int_{0}^{\infty}  d\rho\left[ f_{\rm L}^*\left[\frac{d}{d \rho} \left(\rho \frac{d f_{\rm L}}{d \rho}\right)\right] - \left[ 2i\alpha m \lambda \rho e^{-2\alpha \rho} + \rho e^{-2\alpha \rho} \gamma (r) - k_z^2 \rho  \right]\right] |f_{\rm L}|^2.
\end{equation}

Using the expressions given in (\ref{ax-asy-sols}), we can be sure that
(\ref{lanczos-cond}) can be integrated by parts without adding border
terms. After this procedure one gets: 
\begin{equation}
\int_{0}^{\infty}  d\rho \, \rho \left[\left| \frac{df_{\rm L}}{d\rho}\right|^2 + \left[e^{-2\alpha \rho} \gamma (r) - k_z^2 \right] |f_{\rm L}|^2\right] =  2i\alpha m \lambda \int_{0}^{\infty}  d\rho \, \rho e^{-2\alpha \rho}|f_{\rm L}|^2.\label{vSax}
\end{equation}

As the integrand in the right hand side integral of (\ref{vSax}) is a
positive definite function, we conclude that only axially symmetric
resonant modes ($m=0$) can exist in this spacetime as a real number can not be equal to an imaginary one unless they are both null.

\subsection{The $(2+1)$ extreme BTZ spacetime}

In 1992 M\'aximo Ba\~nados, Claudio Taitelboim and Jorge Zanelli
showed that (2+1)-dimensional gravity admits the black hole solution
known as the BTZ black hole \cite{BTZori}. This solution is not
asymptotically flat, in fact its geometry is asymptotically
AdS. Besides, at the origin, $r=0$, no curvature singularity is
present in the spacetime. The rotating BTZ black hole has two
horizons: an event horizon and an inner Cauchy one. When extended (to
the $r^2<0$ region) BTZ spacetime presents a region with CTC's, a
property shared with Kerr's solution. 

The line element for this spacetime can be written as:
\begin{equation}
ds_{\rm BTZ}^2 = \left(M - \frac{r^2}{l^2}\right) dt^2 - Jdtd\phi + \left(-M +\frac{r^2}{l^2} + \frac{J^2}{4r^2} \right)^{-1}dr^2 + r^2 d\phi ^2.
\end{equation} 

This stationary and axially symmetric spacetime satisfies the vacuum
Einstein field equations with a cosmological constant $\Lambda =
-l^{-2}$. The other two parameters needed to describe its geometry, $M$ and $J$, are respectively the
ADM mass and angular momentum of the spacetime. BTZ spacetime has two
horizons located at: 
\begin{equation}
r^2_{\pm} = \frac{Ml^2}{2}\left[1 \pm \sqrt{1 - \left(\frac{J}{Ml}\right)^2} \right],
\end{equation}

\noindent it also has an ergosphere, whose boundary is located at:
\begin{equation}
r_{\rm erg} = \sqrt{M}l.
\end{equation}

As happens in Kerr's solution, for $r<r_{\rm erg}$ timelike curves
necessarily have (if $J>0$) $d\phi / dt >0$, thus observers suffer the
frame dragging effect. When $|J| > Ml$, the conical singularity
located at $r=0$ becomes naked to distant observers. When $J^2 =
l^2M^2$, the spacetime is called extremal BTZ spacetime.

The stability under linear field perturbations of the non-rotating BTZ
solution was established in Ref.~\refcite{nr-BTZ}. Some aspects of the
stability of the rotating BTZ solution were addressed in
Ref.~\refcite{rev-QNM,r-BTZ,r-BTZ3,r-BTZ2} and the quasi-normal modes
in extreme BTZ studied in Ref.~\refcite{BTZ-ex-qnm}. In what follows,
we study a different family of scalar perturbing
modes. 

It is worth noticing that $l$ can be chosen to be positive with no loss of generality, that $M>0$ is the physically interesting case and that this spacetime is invariant under the simultaneous change $J \to -J$ and $\phi \to -\phi$. Then, we can consider the case $J>0$ with no loss of generality. Thus, the extremal condition can be simplified to: $l = J M^{-1}$. 

For non-axial resonant modes we can describe the perturbation via the
test field $\Phi(r,\phi ,t) = e^{im\phi + \lambda t}f(r)$. Introducing
this expression in the scalar wave equation we get that the radial
part of the perturbation is governed by: 
\begin{equation} \label{btz-11}
\frac{d^2 f(r)}{dr^2} - \frac{J^2 + 6r^2 M}{r(J^2 - 2Mr^2)} \frac{df(r)}{dr} - \frac{16J^2r^2\left[Z(r)- i J^3\lambda m \right]}{(4r^4M^2 - 4J^2Mr^2+J^4)^2}f(r) = 0,
\end{equation}

\noindent where:
\begin{equation}
Z(r) = (J^2 - r^2M)M m^2 - \lambda ^2J^2r^2.
\end{equation}

If we introduce the integrating factor
\begin{equation}
\Lambda^{\rm BTZ_{ex}}(r) = \frac{J^2 - 2r^2M}{\sqrt{r}},
\end{equation}

\noindent and define $F(r) = \Lambda^{\rm BTZ_{\rm ex}}(r)f(r)$, equation (\ref{btz-11}) can be written as:
\begin{equation} \label{btz-intfact}
\frac{d^2 F(r)}{dr^2} - \frac{16J^2r^2}{(4r^4M^2 - 4J^2Mr^2+J^4)^2}\left[Z(r) + \frac{3}{4r^2} - i J^3\lambda m \right]F(r) = 0.
\end{equation}

As we mentioned before, we will analyze three different regions: the interior, $0 < r < r^{\rm ex}_+ = Ml^2/2$, and the exterior one, $r >
r^{\rm ex}_+$ in this Section and in the next one, the region with CTC's, $r^2<0$.

\subsubsection{Interior region}

Taking expression (\ref{btz-intfact}), multiplying it by $F^*(r)$ and
integrating the result in the radial interval $(0,r^{\rm ex}_+)$ we get
the following expression: 
\begin{equation}
\int_0^{r^{\rm ex}_+}dr \left[F^*\frac{d^2F}{dr^2} - \frac{16J^2r^2}{(4r^4M^2 - 4J^2Mr^2+J^4)^2}\left[ Z(r)+ \frac{3}{4r^2} - i J^3\lambda m \right]|F|^2\right] = 0. 
\end{equation}

We can integrate it by parts to get:
\begin{eqnarray} \label{btz-bache}
\left.F^*\frac{dF}{dr}\right|_{0}^{r^{\rm ex}_+} &-& \int_{0}^{r^{\rm ex}_+}dr \left[ \left|\frac{dF}{dr}\right|^2 + \frac{16J^2r^2}{(4r^4M^2 - 4J^2Mr^2+J^4)^2}\left[ Z(r) + \frac{3}{4r^2}\right]|F|^2 \right] = \nonumber \\
&&  i \lambda m \int_0^{r^{\rm ex}_+}dr \left[ \frac{16J^5r^2}{(4r^4M^2 - 4J^2Mr^2+J^4)^2}  |F|^2\right] . 
\end{eqnarray}

An analysis of the local solutions at the origin and at the horizon, allow
us to assure that there exists spatially well behaved modes for which 
the border term in (\ref{btz-bache}) is null. The integrand of the
right hand side is positive definite and the integrand of the left
hand side is manifestly real, thus, equation (\ref{btz-bache}) states that
a imaginary number must be equal a real one unless $m=0$. We conclude
that scalar, resonant, non-axial modes can not
exist. In this way we have proven that in the interior region of an
extreme BTZ spacetime the only (potentially existing) scalar field
resonances are axial. This result is in complete agreement with
previous ones \cite{bache}.

\subsubsection{Exterior region}

The study the exterior region, $r >
r_+^{\rm ex}$, is completely analogous to the one used previously for the interior
region. 

In this way we can prove that non-axial resonances
do not exist in the exterior region of the extreme BTZ spacetime for scalar fields.

\subsection{Rotating wormhole solutions} \label{rot.wh.sec}

The general line element representing a stationary rotating and
transversable wormhole (see, for example,
Ref.~\refcite{visser-LWH}) can be written as: 
\begin{equation}
ds^2_{\rm rwh} = -T^2(r,\theta) dt^2 + \frac{1}{1-\frac{p(r)}{r}}dr^2 + r^2 B^2(r,\theta) \left[ d\theta ^2 + \sin ^2 \theta [d\phi - \Omega(r,\theta) dt]^2\right].
\end{equation}

The throat, $r= p(r)$, joins together two identical asymptotically
flat regions. 

The asymptotic ($r \to \infty$) constrains to the functions $T(r,\theta)$, $B(r,\theta)$, $p(r)$ and $\Omega(r,\theta)$ are the following:
\begin{equation}
T(r,\theta) \to 1, \quad \frac{p(r)}{r} \to 0, \quad B(r,\theta) \to 1, \quad \Omega(r,\theta) \to 0,
\end{equation}
\noindent another restriction is $r \ge p(r)$, with equality holding only at the throat.

We present two different solutions for rotating wormholes. First, the
rigid rotating wormhole that is characterized by the following functions: 
\begin{equation}
T(r,\theta) = B(r,\theta) = 1, \quad p(r) = \frac{p_0^2}{r}, \quad \Omega(r,\theta) = \Omega _0.
\end{equation}

As a second example, we consider the case in which $\Omega$ is a function
of the radial coordinate: 
\begin{equation}
T(r,\theta) = B(r,\theta) = 1, \quad p(r) = \frac{p_0^2}{r}, \quad \Omega(r,\theta) = \frac{2a}{r^3}. 
\end{equation}

Using the symmetries of the background spacetime, an anzats for the
scalar test field can be written as $\Phi (r,\theta, \phi ,t) =
R(r)S(\theta)e^{im\phi}e^{\lambda t}$. Therefore, the scalar wave equation for the models considered in this work, can be separated into an angular and a radial equation, the first one reads: 
\begin{equation}
\frac{1}{\sin \theta} \frac{d}{d\theta}\left(\sin \theta \frac{d}{d\theta}\right) S(\theta) - \left(\frac{m^2}{\sin ^2 \theta} - E_{l}\right) S(\theta) = 0,
\end{equation}
\noindent which is the differential equation of the associated
Legendre polynomials $P_l^m (\cos \theta)$ (whose eigenvalues, given
by $E_{l} = l(l+1)$, do not depend on the value of $m$). The equation
governing the radial part depends on the form that $\Omega (r)$ takes.

\subsubsection{The rigid rotation wormhole}

For the case in which the rotation parameter is constant, $\Omega =
\Omega _0$, the radial equation reads as: 
\begin{equation} \label{radial-WH0}
\frac{d^2R(r)}{dr^2} + \frac{2r^2 - p_0^2}{r(r^2 - p_0^2)}\frac{dR(r)}{dr} + \left[(\kappa - 2im\lambda \Omega_0)r^2 + E_{l}\right]R(r) = 0, 
\end{equation}
\noindent where:
\begin{equation}
\kappa = m^2 \Omega _0^2 - \lambda ^2.
\end{equation}

Introducing the integrating factor given by:
\begin{equation}
\Lambda _0 (r) = \frac{1}{4}\frac{(2r^2 + p_0^2)p_0^2}{r^2(r-p_0)^2(r+p_0)^2},
\end{equation}
\noindent and defining a new radial function, $F_0 (r) = \Lambda _0 (r) R(r)$, equation (\ref{radial-WH0}) can be written as:
\begin{equation}
\frac{d^2F_0(r)}{dr^2} - \left[ \frac{1}{\Lambda _0 (r)} \frac{d^2\Lambda _0(r)}{dr^2} - \left(\kappa r^2 + E_{l}\right) \right] F_0 (r) =  2im\lambda \Omega_0r^2 F_0(r) .
\end{equation}

After multiplying by $F_0^*(r)$ and integrating in the radial
coordinate we get, after noticing that there are solutions with
behaviors at the throat and for large values of $r$ for which the
integration by parts does not introduce border terms, the expression: 
\begin{equation}
\int_{p_0}^{\infty} dr \left[ \left|\frac{dF_0}{dr}\right|^2 + \left[ \frac{1}{\Lambda _0 (r)}\frac{d^2\Lambda _0(r)}{dr^2}  - \left(\kappa r^2 + E_{l}\right) \right] |F_0|^2 \right] = - 2i m \lambda \Omega _0 \int_{p_0}^{\infty} dr \,r^2|F_0|^2.
\end{equation}

The integrand of the right hand side is positive definite so we can
conclude that the only possible resonances in this family of rotating
wormholes must be axial ($m=0$).

\subsubsection{The $\Omega (r) = 2a r^{-3}$ wormhole}

In this case, the radial equation reads as:
\begin{equation} \label{rad-omegar}
\frac{d^{2}R(r)}{d{r}^{2}}+ \frac{ 2r^2-p_0^2}{r(r^2 - p_0^2)} {\frac {dR(r)}{dr}} +\left( 4\,{\frac {{a}^{2}{m}^{2}}{{r}^{4}}}-{\lambda}^{2}{r
}^{2}+E_{l} \right)R(r) = {\frac {4\,iam\lambda}{r}}R(r).
\end{equation}

Equation (\ref{rad-omegar}) has the same structure than the one
associated to the rigid rotation case (\ref{radial-WH0}), hence we can conclude that non-axial resonances are not
possible in this spacetime.

\section{CTC's and axial resonances}

In this Section we present results that reinforce
the conjecture which relates the presence of CTC's in the spacetime with the existence of axial, unstable solutions to the scalar wave equation. We perform this study in the Lanczos dust cylinder and the extreme BTZ black hole. The rotating wormhole solutions studied Section \ref{rot.wh.sec} do not present CTC's so we do not continue the stability study (see, Ref.~\refcite{rot.wh.pert}, where perturbations under a different family of test scalar fields was performed).

\subsection{The Lanczos dust cylinder}

 For the particular case of the axial mode, wave equation
(\ref{edo-Lanczos}) reduces to: 
\begin{equation}
\frac{d^2f_{\rm L}^{\rm ax}(\rho)}{d\rho ^2} + \frac{1}{\rho}\frac{d f_{\rm L}^{\rm ax}(\rho)}{d\rho} - [K^2 (1- \alpha ^2 \rho ^2 ) e^{-2\alpha \rho} + k_z^2]f_{\rm L}^{\rm ax}(\rho) = 0. \label{Lanczos-0}
\end{equation}

First, we analyze the case in which $k_z=0$. The asymptotic behavior for these modes is: 
\begin{equation}
{}^0f_{\rm L}^{\rm ax} (\rho) \sim C_1 + C_2 \ln \rho ,
\end{equation}
\noindent thus, we conclude that no physically relevant solutions
for this particular case exist.  

To finish our study we analyze the $k_z \ne 0$ modes. In this case, no
analytical solution to the differential equation (\ref{Lanczos-0}) is,
to the knowledge of the authors, available. We decided to use a
shooting-to-an-intermediate-point algorithm to find
solutions. Starting with the local solution at the axis and the
asymptotic one and using a Runge-Kutta-Felberg of order $(4,5)$ integrator
and a posterior interpolation of degree 4 we study if this
spacetime admits resonant modes or not. 

In order to make the numerical approach more efficient we introduce
the adimensional coordinate $x = \alpha \rho$ and define, in a similar
manner, new wave parameters $\tilde{K} = \alpha K$ and $\tilde{k}_z =
\alpha k_z$. 

\begin{figure}[pb]
\centerline{\psfig{file=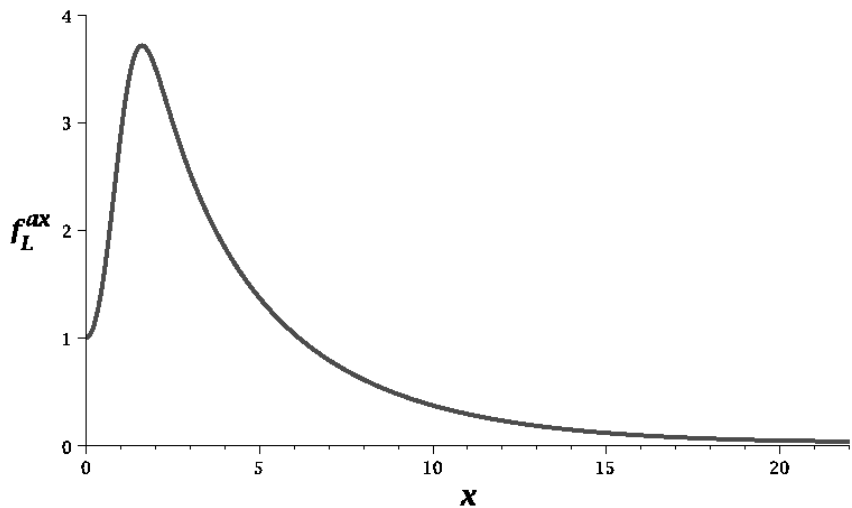,width=11.5cm}}
\vspace*{8pt}
\caption{Typical radial part of a ground state resonant solution to the scalar wave equation in Lanczos spacetime. The resonant mode is characterized by $\tilde{k}_z = 0{.}2$ and $\tilde{K} = 2{.}96327936...$ . The arbitrary amplitude was set to satisfy the condition $f_{\rm L}^{\rm ax}(0) =1$.}
 \label{van02}
\end{figure}

Some of our numerical results are resonant modes characterized by:
\begin{itemize}
\item $\tilde{k}_z = 0{.}2$ and $\tilde{K} = 2{.}96327936...$ (see Figure \ref{van02} for details on this typical solution),
\item $\tilde{k}_z = 2{.}2$ and $\tilde{K}_1 = 8{.}68087136...$ and $\tilde{K}_2 = 15{.}243836...$
\end{itemize}

With our numerical scheme we have covered the range up to $\tilde{k}_z \sim 15$ and have always found resonant modes. These results (although not conclusive) seems to indicate that they exist for all scales.

\subsection{The extreme $(2+1)$ BTZ solution}

\subsubsection{Interior region}

The radial part of the axial mode is governed by the differential equation:
\begin{equation} \label{radial-b}
\frac{d^2 f^{\rm ax}(r)}{dr^2} - \frac{J^2 + 6r^2 M}{r(J^2 - 2Mr^2)} \frac{df^{\rm ax}(r)}{dr} + \frac{16J^4r^4 \lambda ^2}{(4r^4M^2 - 4J^2Mr^2+J^4)^2}f^{\rm ax}(r) = 0.
\end{equation}

For our purposes it is useful to perform the following change in the
radial coordinate: 
\begin{equation}
r^2 =\frac{J^2}{2M}\left(1 - \frac{\sqrt{2}\lambda J}{uM^{3/2}} \right).
\end{equation}

 After this change, equation (\ref{radial-b}) becomes: 
\begin{equation} \label{btz-u}
\frac{d^2 f^{\rm ax}(u)}{du^2} + \frac{1}{4}\left[\frac{u_0}{u} - 1 \right]f^{\rm ax}(u)=0.
\end{equation}

It is possible to see that the $u$-radial coordinate's domain is given
by \hbox{$u_0 \equiv \sqrt{2}\lambda J/M^{3/2} < u < \infty$}. Where
$u \to u_0$ corresponds to $r \to 0$ and $u \to \infty$ to $r \to
r^{\rm ex}_+$. The singular points of equation (\ref{btz-u}) are located at:
\begin{equation}
u = 0 , \qquad u = u_0, \qquad u = \pm \infty ,
\end{equation}
\noindent and its general solution is given by:
\begin{equation} \label{Wsols}
f^{\rm ax} (u) = A_{\rm M} W_{\rm M} \left(\frac{u_0}{4} , \frac{1}{2} ,u \right) + A_{\rm W} W_{\rm W} \left(\frac{u_0}{4} , \frac{1}{2} , u \right),
\end{equation}
\noindent where $W_{\rm M,W}(\mu,\nu,z)$ are the Whittaker functions
and $A_{\rm M,W}$ constants (for details see, for example,
Ref.~\refcite{NIST}).  

Lets review some of their properties. When the first parameter of
Whittaker's functions is not a positive integer, then $W_{\rm M}(a ,
1/2 , u)$ grows exponentially as \hbox{$u \to \infty$}. On the
contrary, $W_{\rm W}(a,1/2,u)$ is well behaved when $u \ge u_0$. 

Equation (\ref{btz-u}) is a time independent Schr\"odinger like
equation with a potential given by $V(u) = -\frac{u_0}{4u}$ and an
eigenenergy $E = -1/4$. As $u=u_0$ corresponds to the classical return
point, we can assure that the eigenfunction $f^0(u)$ would not have
zeros when $u \ge u_0$. Therefore we can prove that properly behaved resonant scalar mode do not exist when $u_0/4$ is not a positive integer.

In the case where $u_0/4 = n$ and $n$
are positive integer numbers, $W_{\rm M}(n,1/2,u)$ and
$W_{\rm W}(n,1/2,u)$ are not linearly independent solutions to
equation (\ref{btz-u}). Using analytical continuation it is possible to construct a second independent solution that is complex-valued, lacking of physical interest and, therefore, not taken into account. To discard
the relevance of the other solution, $W_{\rm W}(n,1/2,u)$, we use the
non existence nodes outside the ``classical region'' of solutions to time independent Schr\"odinger like equations. In this way we can ensure that properly behaved modes do not exist.

\subsubsection{Exterior region}

The only difference with the previous case is that now the domain of
the coordinate $u$ is $-\infty < u < 0$. The differential equation for
the $m=0$-mode does not change compared with the one for
the interior region, so the solutions are the same.  

For negative values of $u$, $W_{\rm W}(a,1/2,u)$ is complex-valued
unless the parameter $a$ is an integer number, when it presents a
divergent behavior as $u \to -\infty$. In the negative real axis, $W_{\rm M}(a,1/2,u)$ is real for all values of the parameter $a$, but
diverges exponentially as $u \to -\infty$, so they are not physically
acceptable. Functions $W_{\rm W}(a,1/2,u)$ and $W_{\rm M}(a,1/2,u)$
are not linearly independent when the parameter $a$ is a natural
number. In that case, the second independent solution can be found
using an analytical extension procedure. The result is that
one of the solutions diverges exponentially as $u \to -\infty$ and the
other does not have a proper limit when $u \to 0$. 

As an example, we present the case $n=3$ in which the two independent
solutions are: 
\begin{eqnarray}
f_1(u) = {{e}^{-\frac{u}{2}}}u \left( 6-6\,u+{u}^{2} \right), \;
f_2(u) = {\rm Ei}\left( 1,-u \right) f_1(u)+ \left( 2-5\,u+{u}^{2} \right) {e}^{\frac{u}{2}}, 
\end{eqnarray}
where ${\rm Ei}(1,-u)$ is the exponential integral function.

As can be seen, $f_1(u)$ diverges for large values of $-u$ and $\lim_{u \to 0}f_2(u) \ne 0$.

Summing up, we have proven that in the exterior of an extreme BTZ black
hole, spatially well behaved resonant modes are not possible. Then,
this spacetime is linearly stable against this family of scalar
perturbations.

\subsubsection{The $r^2 < 0$ region of the $(2+1)$ extreme BTZ solution}

In Ref.~\refcite{2+1}, the authors argue that the region $r^2<0$ must
be cut off from the spacetime. The argument used is that the presence
of a field, like the electromagnetic one, gives rise to a singularity
at $r=0$. 

We study the behavior of solutions to the scalar wave equation in this
region of the extreme BTZ spacetime where CTC's are present, focusing on the existence of axial resonant
modes. Equation (\ref{btz-u}) governs the radial part of the axial
perturbation field. The radial coordinate is in the range $0 < u \le
u_0$, where $u \to u_0$ corresponds to $r^2 \to 0^-$ and $u \to 0$ to
$r^2 \to -\infty$. 

In general, the linearly independent solutions are the Whittaker functions (see equation \ref{Wsols}). Lets analyze the
behavior of $W_{\rm M}(u_0/4,1/2,u)$: at the origin, the Taylor expansion gives: 
\begin{equation}
W_{\rm M}(u_0/4,1/2,u) \sim u -\frac{1}{8}u_0 u^2 + \mathcal{O}(u^3) ,
\end{equation}

\noindent which is properly behaved at the singular point $u=0$. A similar analysis for $u \sim u_0$ gives: 
\begin{equation}
W_{\rm M}(u_0/4,1/2,u) \sim u_0e^{-u_0/2}{}_1{\rm F}_1([1-u_0/4],[2],u_0) + \mathcal{O}(u-u_0),
\end{equation} 
\noindent where ${}_1{\rm F}_1([1-u_0/4],[2],u_0)$ is Kummer's confluent
function that, as a function of $u_0$, has an infinite number of
zeroes, $u^{(k)}_0$. 

Then, if we select $\lambda$ from the family:
\begin{equation}
\lambda ^{(k)} = \frac{M^{3/2}}{\sqrt{2} J} u^{(k)}_0, \:\: k = 1,2, \dots
\end{equation}
\noindent we obtain spatially we behaved, exponentially growing
in time modes. 

We have proven that the $r^2 <0$ region of extreme BTZ is linearly
unstable against scalar perturbations. This result is in agreement
with the one found in Ref.~\refcite{doglers} in Kerr's spacetime and links the existence
of CTC's with the unstable nature of a given spacetime. Moreover, this result
can be used as an additional argument to discard this region from the
extreme BTZ spacetime.

\section{Conclusions}

We have analyzed the scalar wave equation in axially symmetric
spacetimes, some of which possess CTC's. 

We have proved a theorem that states that non-axial scalar resonances
can not exist neither in the extreme BTZ black hole, nor in the
Lanczos dust cylinder nor in the two simple rotating wormhole
solutions (the rigid rotation and the \hbox{$\Omega (r) = 2a r^{-3}$}
cases) that we studied. This results are in complete agreement with
those of Ref.~\refcite{bache}.  

Moreover, we have proved the existence of unstable scalar modes both for the $r^2<0$ region of the extreme BTZ black hole
solution and for the Lanczos dust cylinder. These results can be used as
arguments that favor the proposed relationship between existence of
CTC's and unstable nature of a given spacetime presented in Ref.~\refcite{doglers} for Kerr's spacetime. G\"odel's universe is known to be stable against linear scalar field perturbations \cite{scalar-pert-godel}. It then posses a counterexample to this relationship that could be related to the extremely symmetric nature of this spacetime. Deeper analysis on this matter would be left for future investigations. 

The novel result regarding the unstable nature of the $r^2<0$ region
of extreme BTZ spacetime adds to the series of arguments used to rule
out the physical relevance of this part of the maximally extended
spacetime.

\section*{Acknowledgments}

We would like to thank Prof. Dr. Gustavo Dotti for useful comments and
encouragement during the early stages of this investigation. HV and
IFRS acknowledge support from Universidad Nacional de La Plata. IFRS
is a fellow of CONICET.

\end{document}